# Bulk $NdNiO_2$ is thermodynamically unstable with respect to decomposition while hydrogenation reduces the instability and transforms it from metal to insulator

Oleksandr I. Malyi[1], Julien Varignon[2], and Alex Zunger[1,*]

[1]Renewable and Sustainable Energy Institute, University of Colorado, Boulder, Colorado 80309, USA

[2]Laboratoire CRISMAT, CNRS UMR 6508, ENSICAEN, Normandie Université, 6 Boulevard Maréchal Juin, F-14050 Caen Cedex 4, France

**Email:** Alex.Zunger@colorado.edu

The quest for a Ni-based oxide analog to cuprate $Cu^{2+}(d^9)$ superconductors was long known to require a reduced form $Ni^{1+}$ ($d^9$) as in $A^{3+}Ni^{1+}O_2$, being an extreme oxygen-poor form of the usual $A^{3+}Ni^{3+}O_3$ compound. Through $CaH_2$ chemical reduction of a parent $R^{3+}Ni^{3+}O_3$ perovskite form, superconductivity was recently achieved in Sr-doped $NdNiO_2$ on $SrTiO_3$ substrate. Using density functional theory (DFT) calculations, we find that stoichiometric $NdNiO_2$ is significantly unstable with respect to decomposition into 1/2[$Nd_2O_3$ + NiO + Ni] with exothermic decomposition energy of +176 meV/atom, a considerably higher instability than that for common ternary oxides. This poses the question if the *stoichiometric* $NdNiO_2$ nickelate compound used extensively to model the electronic band structure of Ni-based oxide analog to cuprates and found to be metallic is the right model for this purpose. To examine this, we study via DFT the role of the common H impurity expected to be present in the process of chemical reduction needed to obtain $NdNiO_2$. We find that H can be incorporated *exothermically, i.e., spontaneously* in $NdNiO_2$, even from $H_2$ gas. In the concentrated limit, such impurities can result in the formation of a hydride compound $NdNiO_2H$, which has significantly reduced instability relative to hydrogen-free $NdNiO_2$ (decomposition energy of +80 meV/atom instead of +176 meV/atom). Interestingly, the hydrogenated form has a similar lattice constant as the pure form (leading to comparable X-ray diffraction patterns), but unlike the metallic character of $NdNiO_2$, the hydrogenated form is predicted to be a wide gap insulator thus, requiring doping to create a metallic or superconducting state, just like cuprates, but unlike unhydrogenated nickelates. While it is possible that hydrogen would be eventually desorbed, the calculation suggests that pristine $NdNiO_2$ is hydrogen-stabilized. One must exercise caution with theories predicting new physics in pristine stoichiometric $NdNiO_2$ as it might be an unrealizable compound. Experimental examination of the composition of real $NdNiO_2$ superconductors and the effect of hydrogen on the superconductivity is called for.

## I. Introduction

The recent observation of superconductivity in Sr-doped $NdNiO_2$ grown on $SrTiO_3$ substrate [1] raised hopes for a new paradigm relative to cuprate superconductivity, but along with it also posed questions about the importance of *intrinsic* factors (the basic chemical constitution and bonding in nickelates vs. cuprates), as opposed to *extrinsic* factors (the role of doping, defects, or non-stoichiometry) in comparing the two systems. Interest in nickelates as a comparative paradigm to cuprates has been based on the ability to chemically reduce the stable $A^{3+}Ni^{3+}(\mathbf{d^7})O_3$ compound to $A^{3+}Ni^{1+}(\mathbf{d^9})O_2$, comparing the latter to the isoelectronic $A_2^{3+}Cu^{2+}(\mathbf{d^9})O_4$ compound. The literature[2-13] electronic structure calculations on $NdNiO_2$ were consequently based on the ideal stoichiometric and pristine $P_4/mmm$ $NdNiO_2$ structure with the $Ni^{1+}(d^9)$ ion. These calculations, whether using density functional[2-10] or adding dynamical correlation[6,10-13] found for nonmagnetic [2,3,9,10], or antiferromagnetic/ferromagnetic[4-8], or paramagnetic[6,10-12] $NdNiO_2$ a *metallic ground state*, revealing the first difference with respect to cuprates: the former requires doping to become metallic, unlike the nickelates. Such theoretical investigations of a metallic $NdNiO_2$ phase formed the basis for predicting the role of $NdNiO_2/SrTiO_3$ interface in the superconducting properties[3] and the effect of Sr doping[7]. However, recent experimental advances demonstrate that undoped freestanding $NdNiO_2$ has insulator-like temperature-dependent resistivity[14]. Moreover, recent studies indicate that $NdNiO_2$ samples are often non-stochiometric, for instance, nickel-deficient $NdNi_{1-x}O_2$[14], O-rich $NdNiO_{2+y}$[15], and show the coexistence of different phases such as Ni metal and $NdNiO_2$[16] or $NdNiO_2$ and $NdNiO_3$[17]. These results thus raise the question if pristine stoichiometric $NdNiO_2$ can be used as a model for $d^9$ superconductivity.

DFT is able to assess the stability of a compound with respect to decomposition into competing phases by using the convex hull construct, whereby the lowest formation energy at each composition is contrasted with the formation energies of combinations of competing phases with equivalent composition. Compounds located above the convex hull by an amount $\varepsilon$ are predicted to have exothermic decomposition energy $\varepsilon$. Recognizing that compounds with small metastability energy may still be made as potentially long-lived metastable phases, one generally considers finite but small metastability as potentially realizable [18]. Motivated by this, herein, we study the stability of $NdNiO_2$ with respect to decomposition to competing phases. We find that pure stoichiometric $NdNiO_2$ is a highly unstable phase with the decomposition energy of +176 meV/atom (or +704 meV/formula unit (f.u.)) with respect to $1/2[Nd_2O_3 + NiO + Ni]$, questioning if stoichiometric and pristine $NdNiO_2$ compound is ever realized. Our analysis of 3262 experimentally observed ternary oxides (see Supplemental Material[19]) listed in Materials Project[20] demonstrates (see Fig. 1a) that the vast majority of experimentally observed compounds have DFT decomposition energy $\varepsilon$ less than 100 meV/atom above the calculated convex hull[18], so it looks that pristine $NdNiO_2$ is somewhat extra unstable.

Given the distinguishing feature of synthesis of $NdNiO_2$ via *chemical* reduction of $NdNiO_3$ involving $CaH_2$ or NaH [1,14-16,21-24], and inspired by results of Si *et a*l. on H impurities in $ANiO_2$[25], we undertook a broader stability analysis to include the effect of H incorporation into $NdNiO_2$. We find that hydrogen *spontaneously* incorporated in $NdNiO_2$, resulting in the formation of the hydrogenated and insulating $NdNiO_2H$ compound. The resulting compound is significantly more stable with respect to decomposition than pure $NdNiO_2$ and has similar lattice constants (comparable XRD) to $NdNiO_2$. These results (i) suggest that the experimentally $NdNiO_2$ might have a large, hopefully, the detectable concentration of H impurity, which could affect its properties significantly. To the extent that this is so, (ii) this would point to the importance of considering extrinsic factors (such as spontaneous impurity incorporation [26,27]) in assessing the relevant electronic structure of such compounds; (iii) caution theoreticians from the prediction of new physics and superconductivity mechanisms based on the potentially unrealizable base metallic $NdNiO_2$ compound.

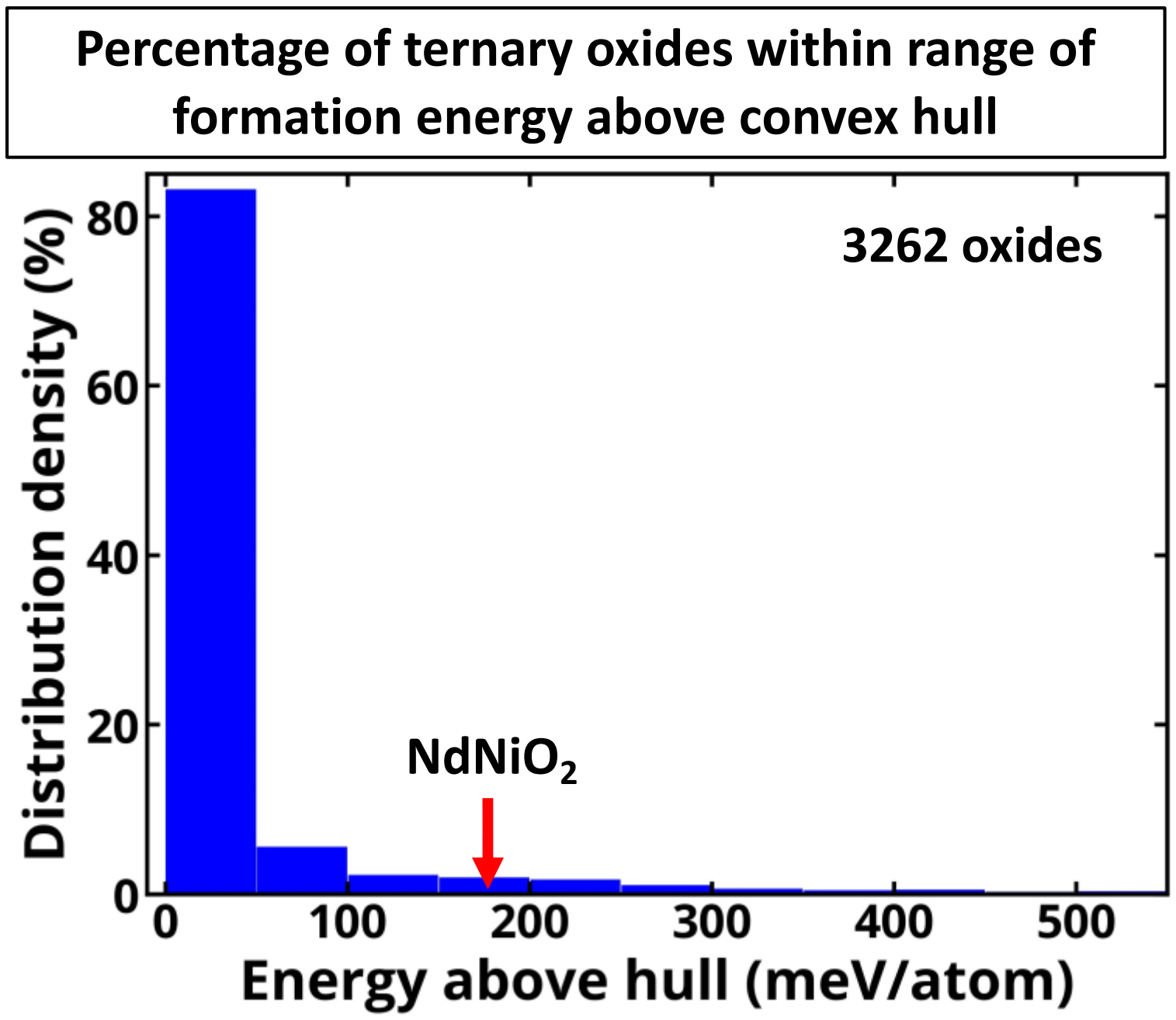

**Figure 1.** Distribution of energy above the DFT convex hull for 3262 experimentally reported ternary oxides (see details on a dataset in Supplemental Material[19]) using the data available in the Materials Project database (accessed on March 20, 2021; Note that numbers available in Materials Project are changing with time)[20]. The red arrow depicts the energy above the convex hull predicted for $NdNiO_2$ ($P_4$/mmm) in this work using accurate DFT with SCAN functional (i.e., one able to accurately treat localized nature of Ni-*d* states) calculations and global screening of magnetic orders in competing phases.

## II. Methods

The calculations are performed using pseudopotential plane-wave DFT with recently developed SCAN exchange-correlation (XC) functional[28] as implemented in the Vienna Ab Initio Simulation Package (VASP)[29-31]. SCAN does not require the Hubbard-like U parameter to reproduce finite band gaps in both AFM and PM binary and ternary oxides[32-35] and has been shown to describe the formation heats of compounds with higher accuracy than other XC functionals[36]. In all calculations, we treat the f electrons of Nd as part of the core electrons. The cutoff energies for the plane-wave basis are set to 500 eV for final calculations and 550 eV for volume relaxation. Atomic relaxations are performed until the internal forces are smaller than 0.01 eV/Å. Γ-centered Monkhorst-Pack grids[37] with 10,000 k-points per reciprocal atom are used in all calculations. For compounds containing less than 9 metal atoms per primitive cell, we explore ferromagnetic, nonmagnetic, and all antiferromagnetic spin configurations that uniquely decorate cells containing 1 or 2 f.u. For compounds constraining more than 8 metal atoms, only ferromagnetic spin initialization has been used. For $NdNiO_2$ and $NdNiO_2H$, all unique antiferromagnetic spin configurations in the supercells up to 4 f.u. are tested. For the compounds containing over 8 atoms, only ferromagnetic and nonmagnetic spin configurations are used. For each structure and magnetic order, we also apply random atomic displacements with a maximum amplitude of 0.1 Å to ensure that the system is not trapped at local energy minima. The computed results are visualized using the Vesta[38] and pymatgen library [39].

## III. Results and Discussion

### A. $NdNiO_2$ is substantially unstable with respect to decomposition: calculation of the ternary convex hull

Using the $O_2$ molecule as a reference state for oxygen and stoichiometric compounds available in Inorganic Crystal Structure Database (ICSD)[40] and Materials Project database[20], we consider 105 different competing phases listed in Supplemental Material[19]. By calculating DFT total energies for these compounds, we obtain the convex hull indicating which Nd-Ni-O compounds are stable (residing on the convex hull) with respect to any linear combination of competing phases or unstable by a given amount (placed above the convex hull by that amount). In this way, 15 out of 105 potential competing phases are found to be stable ground states: $O_2$, Ni, Nd, NiO, $Ni_3O_4$, $NdO_2$, $Nd_2O_3$, NdNi, $NdNi_2$, $Nd_3Ni$, $NdNi_5$, $Nd_5Ni_{19}$, $NdNi_3$, $Nd_2Ni_7$, and $NdNiO_3$. The other competing phases are above the convex hull with the lowest energy decomposition reactions summarized in Table I. Notably, even the lowest energy magnetic $NdNiO_2$ configuration is above the convex hull (unstable) by +176 meV/atom or +704 meV/f.u. with respect to decomposition to 0.5[$Nd_2O_3$ + NiO+ Ni(m)]. There is also a number of other decomposition reactions that have lower decomposition energies (see Fig. 2b). While some oxides that are above the convex hull can be realized, as shown in Fig. 1, over **92%** of experimentally observed oxides have the decomposition energy below that of $NdNiO_2$ according to the data available in the Materials Project. It should also be noted the ICSD database, which is used in Materials Project for experimental entries, has a number of compounds that are predicted by epitaxial growth, observed only under extreme conditions, and compounds that were shown to be incorrect after more detailed analysis of structural properties. Hence, the high decomposition energy of $NdNiO_2$ suggests that stoichiometric $NdNiO_2$ is a rather unstable phase, raising the question if the compound realized experimentally is indeed undoped pristine stoichiometric $NdNiO_2$.

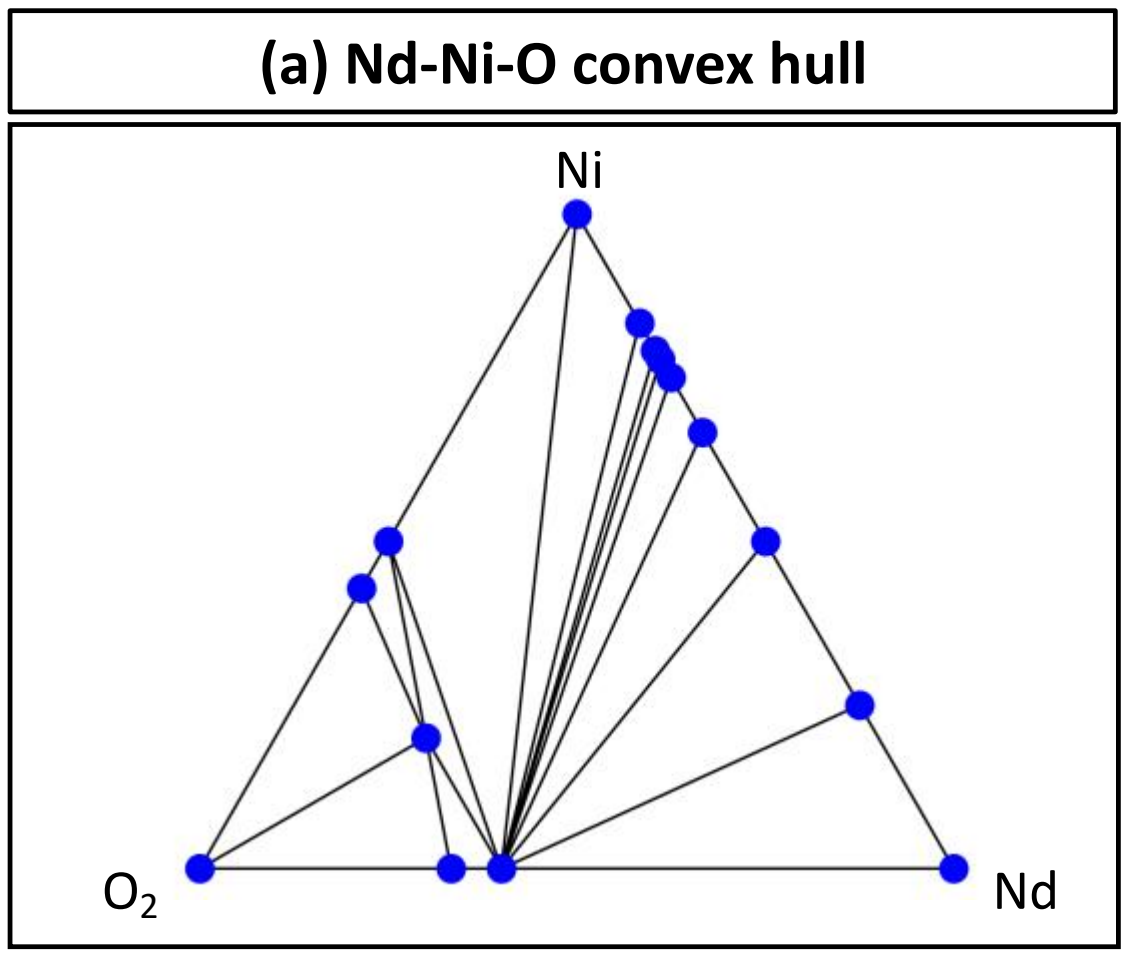

**(b) $NdNiO_2$ main decomposition reactions**

| | |
|---|---|
| $NdNiO_2$ -> 0.111 $NdNi_3$ + 0.444 $Nd_2O$ + 0.667 NiO | $DH_d$ = 62 meV/atom |
| $NdNiO_2$ -> 0.05 $Nd_2Ni_7$ + 0.45 $Nd_2O$ + 0.65 NiO | $DH_d$ = 76 meV/atom |
| $NdNiO_2$ -> 0.019 $Nd_5Ni_{19}$ + 0.453 $Nd_2O$ + 0.642 NiO | $DH_d$ = 83 meV/atom |
| $NdNiO_2$ -> 0.077 $NdNi_5$ + 0.462 $Nd_2O$ + 0.615 NiO | $DH_d$ = 104 meV/atom |
| $NdNiO_2$ -> 0.625 Ni + 0.125 $Ni_3O_4$ + 0.5 $Nd_2O$ | $DH_d$ = 113 meV/atom |
| $NdNiO_2$ -> 0.333 $NdNiO_3$ + 0.667 Ni + 0.333 $Nd_2O_3$ | $DH_d$ = 134 meV/atom |
| $NdNiO_2$ -> 0.5 Ni + 0.5 NiO + 0.5 $Nd_2O_3$ | $DH_d$ = 176 meV/atom |

**Figure 2.** Summary of Nd-Ni-O convex hull. (a) Nd-Ni-O convex hull showing the stable phases ($O_2$, Ni, Nd, NiO, $Ni_3O_4$, $NdO_2$, $Nd_2O_3$, NdNi, $NdNi_2$, $Nd_3Ni$, $NdNi_5$, $Nd_5Ni_{19}$, $NdNi_3$, $Nd_2Ni_7$, and $NdNiO_3$) as blue dots. (b) Main decomposition reactions for $NdNiO_2$ and corresponding decomposition energies ($\Delta H_d$). The results are presented for the SCAN XC functional. The decomposition energies for other compounds are shown in Table I.

**Table I.** Summary of Nd-Ni-O convex hull results including absolutely stable (on hull compounds that do not decompose) as well as metastable (above hull) compounds. Convex hull includes only the lowest energy structure for each composition.

| Compound | Space group | Energy above convex hull (meV/atom) | Decomposition products |
|---|---|---|---|
| Ni | Fm-3m | 0 | - |
| Nd | $P6_3/mmc$ | 0 | - |
| $O_2$ | molecule | 0 | - |
| NiO | C2 | 0 | - |
| $Ni_3O_4$ | Cmmm | 0 | - |
| $Nd_2O_3$ | Ia-3 | 0 | - |
| $NdO_2$ | C2/m | 0 | - |
| NdNi | Cmcm | 0 | - |
| $NdNi_2$ | P1 | 0 | - |
| $NdNi_3$ | R-3m | 0 | - |
| $NdNi_5$ | P6/mmm | 0 | - |
| $Nd_3Ni$ | Pnma | 0 | - |
| $Nd_5Ni_{19}$ | $P6_3/mmc$ | 0 | - |
| $Nd_2Ni_7$ | $P6_3/mmc$ | 0 | - |
| $NdNiO_3$ | $P2_1/c$ | 0 | - |
| $Nd_4Ni_3O_{10}$ | $P2_1/c$ | 1 | $NdNiO_3$, $Nd_2O_3$, NiO |
| $Nd_7Ni_3$ | $P6_3mc$ | 3 | $Nd_3Ni$, NdNi |
| $NiO_2$ | C2/m | 4 | $Ni_3O_4$, $O_2$ |
| $Nd_2Ni_{17}$ | $P6_3/mmc$ | 17 | $NdNi_5$, Ni |
| $Nd_2NiO_4$ | Cmce | 19 | $Nd_2O_3$, NiO |
| NdO | $P6_3/mmc$ | 62 | $Nd_2O_3$, Nd |
| $Ni_5O_6$ | C2/m | 64 | $Ni_3O_4$, NiO |
| $Ni_6O_7$ | P-1 | 70 | $Ni_3O_4$, NiO |
| $Ni_{15}O_{16}$ | Im-3m | 72 | $Ni_3O_4$, NiO |
| $Ni_9O_{10}$ | P-1 | 73 | $Ni_3O_4$, NiO |
| $Nd_4Ni_3O_8$ | I4/mmm | 86 | $Nd_2O_3$, Ni, NiO |
| $Nd_2O_5$ | C2/c | 92 | $NdO_2$, O2 |
| $Ni_5O_{11}$ | P1 | 96 | $Ni_3O_4$, O2 |
| $Ni_5O_4$ | P1 | 165 | Ni, NiO |
| $NdNiO_2$ | P4/mmm | 176 | $Nd_2O_3$, Ni, NiO |
| $NdO_3$ | P-3 | 285 | $NdO_2$, $O_2$ |
| $Nd_4Ni$ | Fd-3m | 364 | $Nd_3Ni$, Nd |

## B. Hydrogenation diminishes the instability of $NdNiO_2$

**Hydrogen as a common additive solute element in $NdNiO_2$ and its bonding to the lattice:** Hydrogen is one of the most common solutes in insulating solids, which often defines the n- and p-type nature of the compounds even under normal conditions owing to universal pinning of H transition level in insulators[41-44]. The most famous examples of such behaviors are ZnO and $SnO_2$, which are intrinsic n-type insulators with hydrogen impurities limiting the possibility to realize p-type insulators [41-44]. In general, hydrogen bonds to anion in n-type compounds (*e.g.*, H-O bond in ZnO[43]) and to cation in p-type compounds[42,45]. In specific cases, hydrogen forms multicenter bonds such as $H_O$ in MgO or ZnO [45], where instead of forming a single bond, it becomes a center of a complex multiple atomic bond inducing charge density rearrangement on multiple atoms. Since $NdNiO_2$ is usually synthesized by the reduction of the stable parent $NdNiO_3$ perovskite in the H-rich environment[1,14-16,21-23], hydrogen impurities can be formed in $NdNiO_2$. To verify it and identify the lowest energy sites, we screen over 35 randomly generated interstitial H positions in $NdNiO_2$ (for simplicity, the screening calculations are performed using PBE XC functional) and find that in the lowest energy configuration H bonds to Ni atoms forming two Ni-H bonds with a bond length of 1.58 Å (Fig. 3a). In such configuration, H accepts electrons from the Ni atoms as evidenced by the charge density difference plot (see Fig. 3a,b), $\Delta\rho = \rho(NdNiO_2{:}H) - \rho(NdNiO_2) - \rho(H)$, where H refers to atomic hydrogen (spin-polarized) and each charge density $\rho$ is computed for the lowest energy configuration of H in 32 f.u. $NdNiO_2$ supercell. These results imply that H acts as the acceptor, restoring part of the charge imbalance caused by the chemical reduction of $A^{3+}Ni^{3+}(\mathbf{d^7})O_3$ compound to $A^{3+}Ni^{1+}(\mathbf{d^9})O_2$.

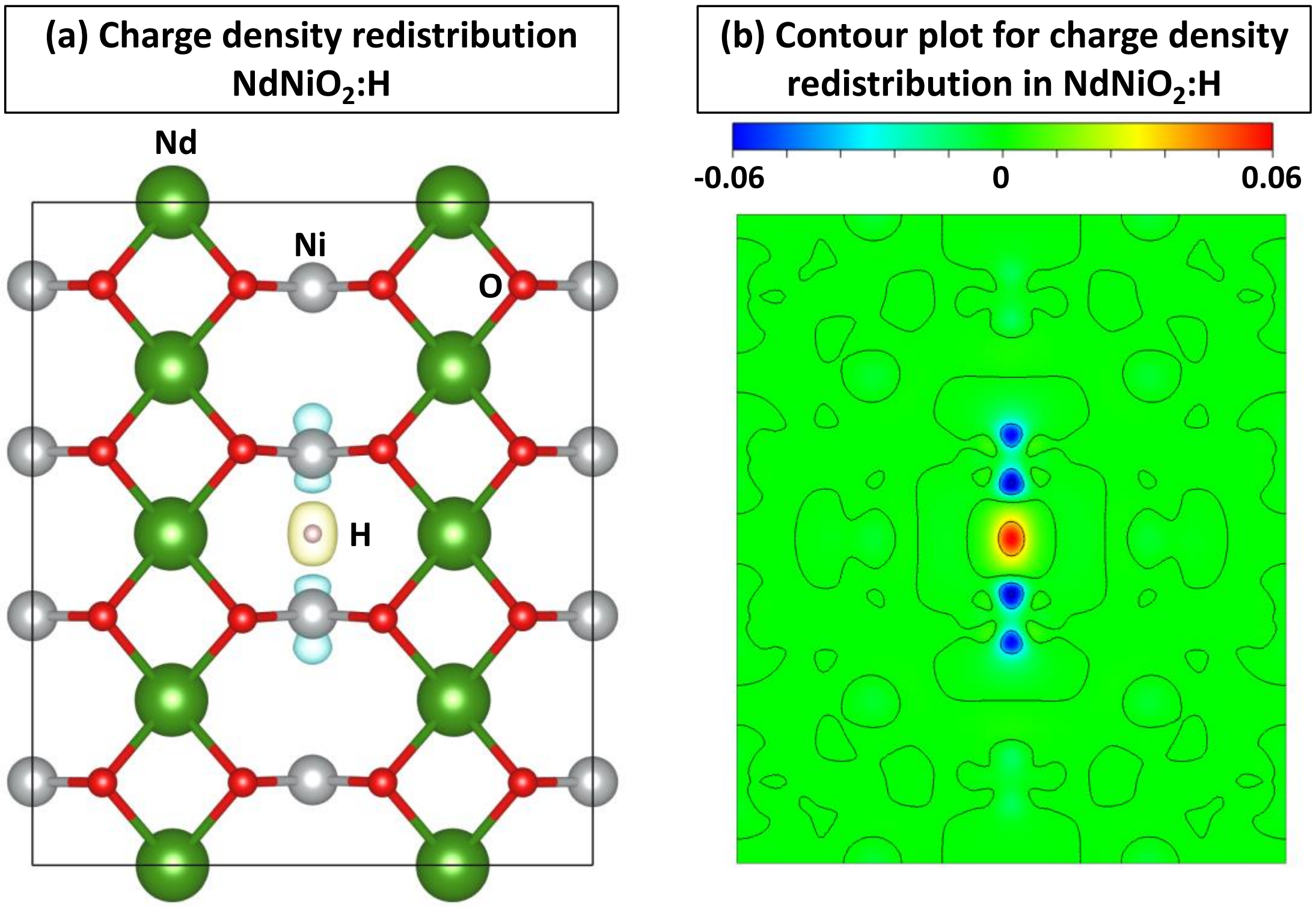

**Figure 3.** 32 f.u. of $NdNiO_2$ containing 1 H impurity: (a) Difference in wavefunction squared – i.e., $NdNiO_2$:H – $NdNiO_2$ -H, where H refers to atomic hydrogen (spin-polarized) and each charge density $\rho$ is computed for lowest energy configuration of H in 32 f.u. $NdNiO_2$ supercell, illustrating the nature of bonding. Yellow and blue isosurfaces correspond to charge density increase and charge density reduction, respectively. The isosurface is set up at 0.01 e/bohr$^3$. Only the atomic planes in the vicinity of H are shown. (b) The corresponding counter-plot for the difference in wavefunction squared shown in (a).

**Spontaneous hydrogen incorporation into $NdNiO_2$**: While it usually costs energy to dissolve hydrogen into traditional oxides[41-44], for $NdNiO_2$, the reaction energy $\Delta H_R$ of H addition to $NdNiO_2$ from $H_2$ gas, *i.e.*,

$$\Delta H_R = E(NdNiO_2{:}H) - E(NdNiO_2) - 1/2H_2(g)$$

is negative (-350 meV), as found here by DFT calculations for 32 f.u. of $NdNiO_2$ containing one interstitial H. This thus suggests that in the presence of an $H_2$-rich environment, H will be spontaneously (exothermically) introduced into $NdNiO_2$. Hence, hydrogen insertion can reduce the instability of $NdNiO_2$-like structure in the limit of high H concentration, i.e., the formation of $(NdNiO_2)_nH_m$ compound can be observed as the result of hydrogen interaction with $NdNiO_2$. To reveal that this type of additive/solute can be formed in large H concentrations, we consider the possibility of the formation of $(NdNiO_2)_nH_m$ with m=n=1 and find that indeed reaction energy of $NdNiO_2 + 1/2\ H_2(g) \rightarrow NdNiO_2H$ is highly negative (-300 meV) (see Fig. 4). These data thus demonstrate that H can be not only a common solute additive but can also lead to the formation of a stoichiometric H-rich $NdNiO_2H$ compound, which is more stable with respect to decomposition than hydrogen free $NdNiO_2$. Importantly, the predicted reaction energy for hydrogen insertion into $NdNiO_2$ is comparable to that in the recent study of H insertion into $ABO_2$[25]. The H insertion changes the stability of $NdNiO_2$ itself. As illustrated in Fig. 4, $NdNiO_2$ decomposes to Ni, NiO, and $Nd_2O_3$ with the decomposition energy of 176 meV/atom, while although $NdNiO_2H$ is still not the ground state – the compound decomposes to Ni, NiO, $Nd_2O_3$, and $H_2$, the corresponding decomposition energy is only 80 meV/atom. While this work does not describe the full range of m/n ratio, the above results intimate that it is likely to observe $NdNiO_2H$ than $NdNiO_2$ especially taking into account that $NdNiO_2$ is usually synthesized by reduction of $NdNiO_3$ in H-rich environment and pure stochiometric $NdNiO_2$ is a highly unstable phase. We note that the formation of $NdNiO_{2.3}H_{0.7}$ has already been detected experimentally[21], suggesting that $NdNiO_2$ indeed can take a large concentration of H defects. It should be noted that herein the results are presented for bulk $NdNiO_2$ and it is possible that in some experimental studies substrate or compering phases (e.g., $SrTiO_3$) can modify $NdNiO_2$ stability and H intake.

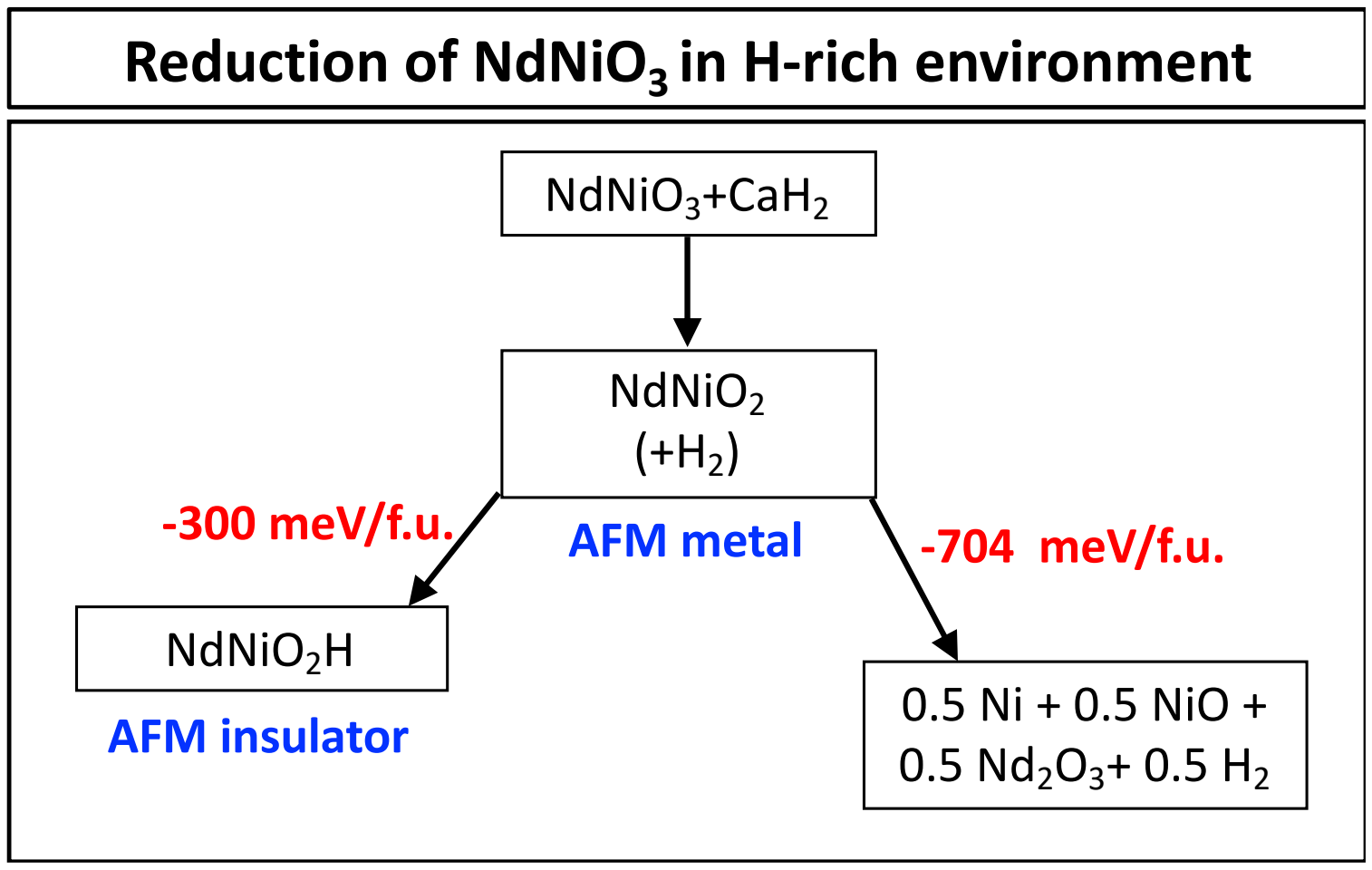

**Figure 4.** Reduction of $NdNiO_3$ and main decomposition reaction for $NdNiO_2$, demonstrating that $NdNiO_2$ is unstable with respect to decomposition to $Nd_2O_3$, NiO, and Ni. Hydrogenation of $NdNiO_2$ reduces the instability of the compound, but even $NdNiO_2H$ is still unstable with respect to competing phases.

**$NdNiO_2H$ and $NdNiO_2$ have comparable lattice constants leading to similar expected X-ray diffraction patterns:** One may think that insertion of large H concentration into $NdNiO_2$ leads to a noticeable structural change leading to a new structure which can be detected in the typical XRD analysis of the sample. However, H is a small solute which usually does not provide significant structural changes. The SCAN lattice constants of $NdNiO_2H$ are 3.95 and 3.35 Å, which are only slightly larger (within the typical difference range of DFT and experimental lattice constants) than the SCAN values for $NdNiO_2$ (3.91 and 3.27 Å) and corresponding experimental lattice constants (e.g., 3.9 and 3.3 Å[21], or 3.91 and 3.24 Å[14], or 3.92 and 3.28 Å[15]). The main XRD peaks for $NdNiO_2$ and $NdNiO_2H$ almost overlap with each other (see Supplemental Material[19]). While a minor difference of lattice vectors between experimental $NdNiO_2$ and theoretical $NdNiO_2$ and $NdNiO_2H$ is reflected in the position of the minor peaks in the XRD spectra as compared to the experimental structure, such peaks are usually not used to identify the compound, and it is well known that the presence of H usually cannot be detected by XRD. If the lattice parameters of $NdNiO_2$ and $NdNiO_2H$ are aligned to experimental lattice constants, both $NdNiO_2$ and $NdNiO_2H$ have the same XRD spectra. Interestingly, $NdNiO_{2.3}H_{0.7}$ and $NdNiO_2$ have a more pronounced difference in XRD spectra[21], which is likely due to deviation from stoichiometry on the O site and not caused by H impurity. These results thus demonstrate that while the formation of $NdNiO_2H$ is spontaneous its detection is not possible with standard XRD and more accurate methods should be applied. Moreover, since $NdNiO_2$ is rarely synthesized in pure form, e.g., experimentally observed $NdNiO_2$ samples are nickel deficient [14], O-rich [15] or have coexistence of different phases [16,17], the determination of exact composition for different samples is critical step in understanding better current experimental observations.

***Possible role of substrate:*** Examining the simplest substrate effect of potential increase in lattice mismatch between the $SrTiO_3$ substrate due to the presence of H in the $NdNiO_2$ film suggest that this factor may not be a concern. Indeed, the DFT calculated in-plane lattice constant of $NdNiO_2$ changes by H incorporation by only 0.03 Å, having a negligible contribution to the altered strain and altered stabilization of a hydrogen-free $NdNiO_2$ on $SrTiO_3$. However, in general, it is essential to further understand the potential role of the substrate and spontaneous off-stoichiometry in the stability of $NdNiO_2$ and H insertion.

## C. Effect of Hydrogen on electronic properties of $NdNiO_2$

**Hypothetical $NdNiO_2$:** Similar to previous DFT[2-10] and dynamical mean-field theory [6,10-13] studies, we find that unstable pristine $NdNiO_2$ (Fig. 5a) is metallic as confirmed by the analysis of the spin-projected density of states (DOS) for the lowest energy magnetic order (Fig. 5c). The DOS has a strong Ni peak located below the Fermi level, and O-p mainly contributes a few eV below it. Importantly, the majority of previous studies[2,3,9,10] utilized a nonmagnetic spin-configuration. Such selection has been justified by paramagnetic (PM) nature of the experimentally observed Sr-doped $NdNiO_2$ compound. However, as has been demonstrated recently, nonmagnetic approximation cannot be used to describe the PM compounds having a set of local non-zero magnetic moments and total magnetic moment zero[35,46-49]. In fact, paramagnets have different local spin environments $\{S_i;i=1, N\}$, and hence then its physical property P (e.g., electronic structure) cannot be approximated as the property $<P>=P(S_0)$ of the macroscopically averaged structure $S_0$, instead of the correct average $P_{obs}=\Sigma P(S_i)$ of the properties $\{P(S_i)\}$ of the individual, low symmetry microscopic configurations [35,46-49]. Moreover, for the pure $NdNiO_2$, the lowest energy AFM state is about 100 meV/atom lower than high energy hypothetical nonmagnetic state according to our SCAN calculations. These results thus intimate that some of the early theoretical predictions based on the nonmagnetic assumption may be misleading. It should also be noted that the AFM nature of pure stoichiometric $NdNiO_2$ has been confirmed by other theoretical works as well[5-8], which also reported a noticeable difference in the electronic properties of nonmagnetic and magnetic $NdNiO_2$.

**$NdNiO_2$:H:** Hydrogenation of $NdNiO_2$ leads to a significant change in electronic properties. For instance, $NdNiO_2H$ is predicted to be an AFM insulator with DFT band gap energy of about 0.7 eV (Fig. 5b,d). While we cannot argue that experimentally reported structures are $NdNiO_2H$, we would like to note that experimentally reported freestanding "$NdNiO_2$" compound is indeed an insulator as confirmed by measurements of electronic conductivity/resistivity vs. temperature[14], which clearly contradicts to theoretical predictions for pure stoichiometric $NdNiO_2$ in this work and that available in the literature[2-13]. These observations thus suggest that before understanding the physics of superconductivity in nickelates, one might first need to understand the composition of samples and the role of common impurities such as H in the stabilization of $NdNiO_2$-like structure or even more complex deviations from ideal stoichiometry. These results also exercise serious caution to predicting new physics in pristine stoichiometric $NdNiO_2$ as it might be a potentially unrealizable compound.

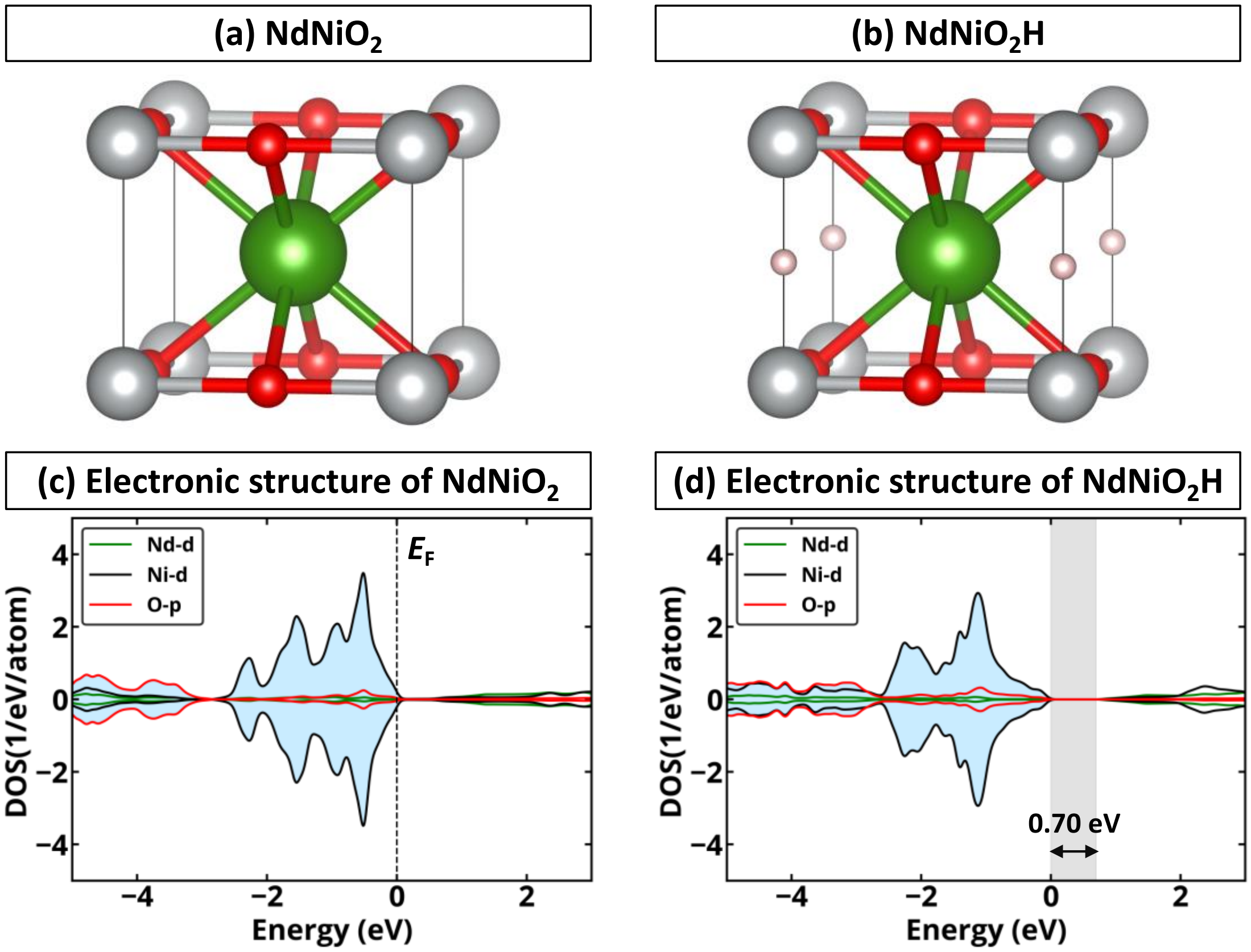

**Figure 5**. Crystal structures of (a) $NdNiO_2$ and (b) hydrogenated $NdNiO_2H$. Orbital projected spin-polarized density of states of pure stoichiometric P4/mmm $NdNiO_2$ (c) and (b) hydrogenated $NdNiO_2$ using the SCAN exchange-correlation functional. Occupied states are shadowed in light blue. The band gap is shown in gray.

## IV. Conclusions

This study demonstrates that pure stoichiometric $NdNiO_2$ is significantly thermodynamically unstable with respect to decomposition into $Nd_2O_3$, NiO, and Ni. Given the distinguishing feature of synthesis of $NdNiO_2$ *via* chemical reduction of $NdNiO_3$ involving $CaH_2$ or NaH, we point out that hydrogen inclusions can form spontaneously in stoichiometric $NdNiO_2$. In the concentrated limit, such inclusions can result in formation of $NdNiO_2H$. Although $NdNiO_2H$ is also unstable with respect to decomposition, its decomposition energy to 0.5[$Nd_2O_3$+NiO+Ni+$H_2$] is only 80 meV/atom, substantially lower compared to the pristine phase. Hydrogenation introduces a fundamental change in the electronic structure, converting the pristine $NdNiO_2$ from AFM metal to a $NdNiO_2$:H AFM insulator, requiring doping as a prerequisite for superconductivity. These results point to the importance of considering extrinsic factors (such as spontaneous defect physics) in assessing the relevant electronic structure of such compounds. ***The need for experimental determination of H incorporation in as grown and subsequent H retention:*** Indeed, there is precedence for incorporation of H as hydride when such reductions are done, as noted in refs. [21,50], in particular, when late row transition metals (e.g., Ni) are present. Such experiments might involve Secondary Ion Mass Spectroscopy (SIMS) or neutron diffraction. This could determine the concentration of hydrogen in the *as prepared* sample, predicted here thermodynamically to bind significant H concentration, and establish if the *processed sample,* used in superconductivity measurement, might have retained hydrogen, or else it escaped. Hence, further understanding the physics of superconductivity in reduced nickelates might require investigation of the role of H impurities or even more complex deviations from ideal stoichiometry (e.g., formation of H-doped $Nd_{1-x}Ni_{1-y}O_2$ compounds). Keeping this in mind, theorists should be aware that pure stoichiometric $NdNiO_2$ may be a "fantasy material"[51] that may never be realized as such, and hence the theoretical predictions based on such material might be misleading.

**Acknowledgment:** The theory of synthesis in this work was supported by Air Force Office of Scientific Research under MURI Award No. FA9550-18-1-0136. The electronic structure and doping calculations were supported by U.S. Department of Energy, Office of Science, Basic Energy Sciences, Materials Sciences and Engineering Division within DE-SC0010467. The authors acknowledge the use of computational resources located at the National Renewable Energy Laboratory and sponsored by the Department of Energy's Office of Energy Efficiency and Renewable Energy. This work also utilized the Extreme Science and Engineering Discovery Environment (XSEDE) supercomputer resources, which are supported by the National Science Foundation grant number ACI-1548562. JV acknowledges access granted to HPC resources of Criann through the project 2020005 and of Cines through the DARI project A0080911453. The authors thank Harold Y. Hwang and Tyrel McQueen for discussions of pertinent experimental aspects.

## References

[1] D. Li, K. Lee, B. Y. Wang, M. Osada, S. Crossley, H. R. Lee, Y. Cui, Y. Hikita, and H. Y. Hwang, Nature, **572**, 624 (2019).
[2] Y. Nomura, M. Hirayama, T. Tadano, Y. Yoshimoto, K. Nakamura, and R. Arita, Phys. Rev. B, **100**, 205138 (2019).
[3] B. Geisler and R. Pentcheva, Phys. Rev. B, **102**, 020502(R) (2020).
[4] P. Jiang, L. Si, Z. Liao, and Z. Zhong, Phys. Rev. B, **100**, 201106 (2019).
[5] M.-Y. Choi, K.-W. Lee, and W. E. Pickett, Phys. Rev. B, **101**, 020503 (2020).
[6] J. Karp, A. S. Botana, M. R. Norman, H. Park, M. Zingl, and A. Millis, Phys. Rev. X, **10**, 021061 (2020).
[7] H. Zhang, L. Jin, S. Wang, B. Xi, X. Shi, F. Ye, and J.-W. Mei, Phys. Rev. Res., **2**, 013214 (2020).
[8] Z. Liu, Z. Ren, W. Zhu, Z. Wang, and J. Yang, npj Quantum Mater., **5**, 1 (2020).
[9] M. Hirayama, T. Tadano, Y. Nomura, and R. Arita, Phys. Rev. B, **101**, 075107 (2020).
[10] F. Lechermann, Phys. Rev. B, **101**, 081110 (2020).
[11] I. Leonov, S. L. Skornyakov, and S. Y. Savrasov, Phys. Rev. B, **101**, 241108 (2020).
[12] I. Leonov and S. Y. Savrasov, arXiv:2006.05295 [cond-mat], (2020).
[13] M. Kitatani, L. Si, O. Janson, R. Arita, Z. Zhong, and K. Held, npj Quantum Mater., **5**, 1 (2020).
[14] Q. Li, C. He, J. Si, X. Zhu, Y. Zhang, and H.-H. Wen, Communications Materials, **1**, 1 (2020).

[15] M. A. Hayward and M. J. Rosseinsky, Solid State Sci., **5**, 839 (2003).
[16] B.-X. Wang, H. Zheng, E. Krivyakina, O. Chmaissem, P. P. Lopes, J. W. Lynn, L. C. Gallington, Y. Ren, S. Rosenkranz, J. F. Mitchell, and D. Phelan, Physical Review Materials **4**, 084409 (2020).
[17] X.-R. Zhou, Z.-X. Feng, P.-X. Qin, H. Yan, S. Hu, H.-X. Guo, X.-N. Wang, H.-J. Wu, X. Zhang, H.-Y. Chen, X.-P. Qiu, and Z.-Q. Liu, Rare Met., **39**, 368 (2020).
[18] W. Sun, S. T. Dacek, S. P. Ong, G. Hautier, A. Jain, W. D. Richards, A. C. Gamst, K. A. Persson, and G. Ceder, Sci. Adv., **2**, e1600225 (2016).
[19] See Supplemental Material for further details on the data used from Materials Project, comparison of XRD spectra for $NdNiO_2$ and $NdNiO_2H$ compounds, and summary on used compounds in the convex hull calculations.
[20] A. Jain, S. P. Ong, G. Hautier, W. Chen, W. D. Richards, S. Dacek, S. Cholia, D. Gunter, D. Skinner, G. Ceder, and K. A. Persson, APL Mat., **1**, 011002 (2013).
[21] T. Onozuka, A. Chikamatsu, T. Katayama, T. Fukumura, and T. Hasegawa, Dalton Trans., **45**, 12114 (2016).
[22] S. Zeng, C. S. Tang, X. Yin, C. Li, Z. Huang, J. Hu, W. Liu, G. J. Omar, H. Jani, Z. S. Lim, K. Han, D. Wan, P. Yang, A. T. S. Wee, and A. Ariando, Phys. Rev. Lett., **125**, 147003 (2020).
[23] K. Lee, B. H. Goodge, D. Li, M. Osada, B. Y. Wang, Y. Cui, L. F. Kourkoutis, and H. Y. Hwang, APL Mat., **8**, 041107 (2020).
[24] D. Li, B. Y. Wang, K. Lee, S. P. Harvey, M. Osada, B. H. Goodge, L. F. Kourkoutis, and H. Y. Hwang, Phys. Rev. Lett., **125**, 027001 (2020).
[25] L. Si, W. Xiao, J. Kaufmann, J. M. Tomczak, Y. Lu, Z. Zhong, and K. Held, Phys. Rev. Lett., **124** 166402 (2020).
[26] A. Walsh and A. Zunger, Nat. Mater., **16**, 964 (2017).
[27] O. I. Malyi, M. T. Yeung, K. R. Poeppelmeier, C. Persson, and A. Zunger, Matter, **1**, 280 (2019).
[28] J. Sun, A. Ruzsinszky, and J. P. Perdew, Phys. Rev. Lett., **115**, 036402 (2015).
[29] G. Kresse and J. Hafner, Phys. Rev. B, **47**, 558 (1993).
[30] G. Kresse and J. Furthmüller, Comp. Mater. Sci., **6**, 15 (1996).
[31] G. Kresse and J. Furthmüller, Phys. Rev. B, **54**, 11169 (1996).
[32] O. I. Malyi and A. Zunger, Appl. Phys. Rev., **7**, 041310 (2020).
[33] Z. Wang, O. I. Malyi, X. Zhao, and A. Zunger, Phys. Rev. B, **103** 165110 (2021).
[34] Y. Zhang, J. Furnes, R. Zhang, Z. Wang, A. Zunger, and J. Sun, Phys. Rev. B, **102**, 045112 (2020).
[35] J. Varignon, M. Bibes, and A. Zunger, Phys. Rev. B, **100**, 035119 (2019).
[36] C. J. Bartel, A. W. Weimer, S. Lany, C. B. Musgrave, and A. M. Holder, npj Comput. Mater., **5**, 4 (2019).
[37] H. J. Monkhorst and J. D. Pack, Phys. Rev. B, **13**, 5188 (1976).
[38] K. Momma and F. Izumi, J. Appl. Crystallogr., **44**, 1272 (2011).
[39] S. P. Ong, W. D. Richards, A. Jain, G. Hautier, M. Kocher, S. Cholia, D. Gunter, V. L. Chevrier, K. A. Persson, and G. Ceder, Comp. Mater. Sci., **68**, 314 (2013).
[40] A. Belsky, M. Hellenbrandt, V. L. Karen, and P. Luksch, Acta Crystallogr., Sect. B: Struct. Sci., **58**, 364 (2002).
[41] C. Kilic and A. Zunger, Appl. Phys. Lett., **81**, 73 (2002).
[42] C. G. Van de Walle and J. Neugebauer, Nature, **423**, 626 (2003).
[43] C. G. Van de Walle, Phys. Rev. Lett., **85**, 1012 (2000).
[44] A. K. Singh, A. Janotti, M. Scheffler, and C. G. Van de Walle, Phys. Rev. Lett., **101**, 055502 (2008).
[45] A. Janotti and C. G. Van de Walle, Nat. Mater., **6**, 44 (2007).
[46] Z. Wang, X. Zhao, R. Koch, S. J. L. Billinge, and A. Zunger, Phys. Rev. B, **102**, 235121 (2020).
[47] G. Trimarchi, Z. Wang, and A. Zunger, Phys. Rev. B, **97**, 035107 (2018).
[48] X.-G. Zhao, G. M. Dalpian, Z. Wang, and A. Zunger, Phys. Rev. B, **101**, 155137 (2020).
[49] J. Varignon, M. Bibes, and A. Zunger, Nat. Commun., **10**, 1658 (2019).
[50] M. Hayward, E. Cussen, J. Claridge, M. Bieringer, M. Rosseinsky, C. Kiely, S. Blundell, I. Marshall, and F. Pratt, Science, **295**, 1882 (2002).
[51] G. A. Sawatzky, Nature, **572**, 592 (2019).